\documentclass[lettersize,journal]{IEEEtran}
\usepackage{amsmath,amsfonts}
\usepackage{algorithmic}
\usepackage{tabularx}
\usepackage{array}
\usepackage[caption=false,font=normalsize,labelfont=sf,textfont=sf]{subfig}
\usepackage{textcomp}
\usepackage{stfloats}
\usepackage{url}
\usepackage{verbatim}
\usepackage{graphicx}
\def\BibTeX{{\rm B\kern-.05em{\sc i\kern-.025em b}\kern-.08em
    T\kern-.1667em\lower.7ex\hbox{E}\kern-.125emX}}
\usepackage{balance}
\usepackage{stfloats}

\usepackage{hyperref}
\hypersetup{
     colorlinks=true,
     linkcolor=blue,
     filecolor=blue,
     citecolor = blue,      
     urlcolor=blue,
     }
     
\usepackage[nameinlink]{cleveref}
\Crefname{figure}{Fig.}{Figs.}
\usepackage{textgreek}
\usepackage{cite}
\usepackage[skip=10pt]{caption}

\usepackage{todonotes}
\usepackage{mwe}
\usepackage{float}
\usepackage{placeins}
\usepackage{upgreek}

\begin{document}
\title{A 40.68-MHz, 200-ns-Settling Active Rectifier and TX-Side Load Monitoring for Minimizing Radiated Power in
Biomedical Implants}
\author{Ronald Wijermars, Yi-Han Ou-Yang, Sijun Du, Dante Gabriel Muratore
\thanks{The authors are with the Department of Microelectronics, Delft University
of Technology, 2628 CD Delft, The Netherlands (e-mail: rgjwijermars@tudelft.nl).}
}

\markboth{IEEE SOLID-STATE CIRCUITS LETTERS, VOL. X, 2025}%
{How to Use the IEEEtran \LaTeX \ Templates}

\maketitle

\begin{abstract}
This letter describes a 40.68~MHz wireless power transfer receiver for implantable applications focused on minimizing tissue heating. The system features a novel power radiated efficiency optimization strategy and a fast-settling active rectifier that maintains high efficiency during load and link variations required for downlink communication. The power radiated efficiency optimization explicitly reduces tissue heating while enabling transmitter-side load monitoring for closed-loop control. The active rectifier was fabricated in 40nm CMOS and achieves a voltage conversion ratio of 93.9\% and a simulated power conversion efficiency of 90.1\% in a 0.19~mm\textsuperscript{2} area, resulting in a 118~mW/mm\textsuperscript{2} power density while integrating the resonance and filter capacitors. The worst-case settling of the on- and off-delay compensation in the active rectifier is 200~ns, which is the fastest reported to date. 
\end{abstract}


\section{Introduction}

Wireless power transfer (WPT) can reduce invasiveness in biomedical implants by eliminating the need for wires or batteries. Notably, these implants are generally limited by the power dissipated in the surrounding tissue, requiring careful design of the power management unit (PMU). Typically, the PMU implements a rectifier followed by a linear regulator to provide a stable DC output. The rectifier stage is typically implemented using comparator-based active structures to achieve high power conversion efficiency (PCE). These active rectifiers require adaptive ON/OFF delay compensation, rather than fixed delay schemes, to maintain optimal PCE under varying input and loading conditions \cite{Lee_2012,Cheng_2016,Huang_2016,Xue_2019,Lu_2014,Ahn_2024}. Adaptive delay compensation is generally accomplished with a feedback loop that consists of a sampling circuit and an error amplifier. The amplifier introduces an offset into a zero-crossing comparator to compensate for the ON/OFF delay. To ensure system stability, the amplifiers are intentionally designed with low bandwidth \cite{Bai_2023,Huang_2016,Luo_2023,Li_2015,Xue_2019,Cheng_2016}. However, this low bandwidth inherently limits their response time during fast input or load variations. This is even more critical in implants, where amplitude shift-keying (ASK) is commonly used to transmit down-link data on the power carrier \cite{Lee_2012}.

At the system level, the PMU is typically optimized for power transfer efficiency (PTE) at the expected system load \cite{Kiani_2022,Ramrakhyani_2011,Li_2015}. In this approach, the optimal PTE is achieved only for a single load scenario, and any deviation in the system load corresponds to a deviation from the optimum. Furthermore, since the radiated power is proportional to the received voltage, the linear regulator in the PMU sets a lower limit on the power that needs to be radiated through the tissue to maintain a sufficiently high voltage for the load. Finally, the optimal PTE load focuses on the input power to the transmitter, and it does not necessarily minimize the radiated power into the tissue, which is the limiting factor for safety.

\begin{figure}[t]
    \centering
    \includegraphics[clip, trim=0.5cm 0.5cm 0cm 0cm,width=0.95\linewidth]{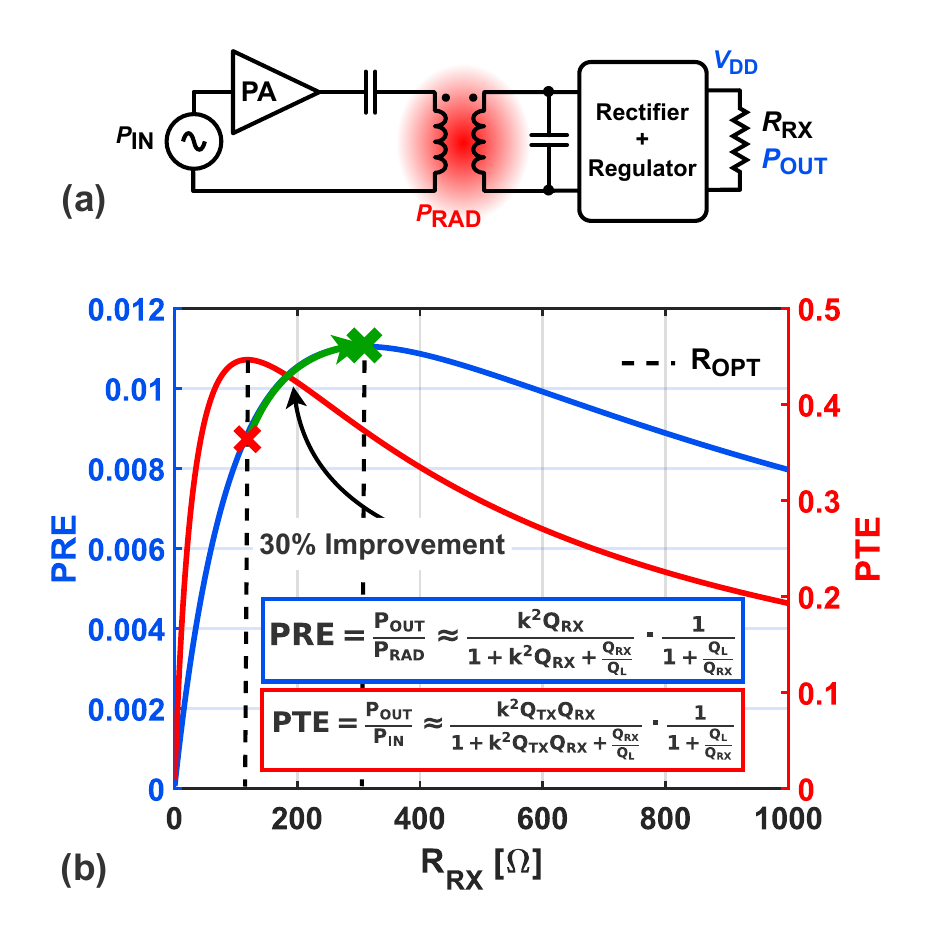}
    \caption{(a) System architecture of WPT-link and PMU, (b) Analytical analysis PRE improvement compared to PTE for an existing link.}
    \label{fig:Conventional_vs_proposed}
\end{figure}

This letter presents two key concepts. (i) We introduce an active receiver (RX) rectifier with rapid ON/OFF adaptive delay compensation for a 40.68~MHz inductive WPT. The rectifier design implements a novel direct voltage-domain compensation and eliminates the need for a slow feedback control loop while maintaining high PCE under fast input or load variations. The high resonance frequency enables on-chip integration of the RX resonance and filter capacitor, requiring only a coil as an external component. A prototype of the fast-settling rectifier achieved a measured VCR of 93.9\% and a simulated PCE of 90.1\% in a 0.19~mm\textsuperscript{2} active area, which includes the resonant and filter capacitors. (ii) We propose a new link optimization method based on power radiated efficiency (PRE) that minimizes radiated power for a given output load. If combined with a switched capacitor power converter (SCPC), this approach simultaneously maintains optimum PRE across load variations and enables TX-side load monitoring to regulate the transmitted power.

\section{Power Radiated Efficiency}

PRE is defined as the ratio between the power radiated by the TX antenna (P\textsubscript{RAD}) and the output power in the RX (\textit{P}\textsubscript{OUT}), while PTE is defined as the ratio between the input power to the TX power amplifier (P\textsubscript{IN}) and the output power in the RX (\textit{P}\textsubscript{OUT}). Since the TX is typically outside the body, its power efficiency is less critical and the main goal is to reduce the power radiated into the tissue. Hence, the objective is to maximize PRE and not PTE. A potential reduction of 30\% in radiated power can be achieved in our WPT link (details in Section \ref{sec-meas}) by optimizing the RX tank load (\textit{R}\textsubscript{RX}) for PRE rather than PTE, assuming no additional losses  (\Cref{fig:Conventional_vs_proposed}(b)). This improvement occurs because the TX power sent to the resonant tank (\textit{P}\textsubscript{IN}) is much smaller than the power in the tank itself, since most of the power is reactive. 

\begin{figure}[t]
    \centering
    \includegraphics[width=0.9\linewidth]{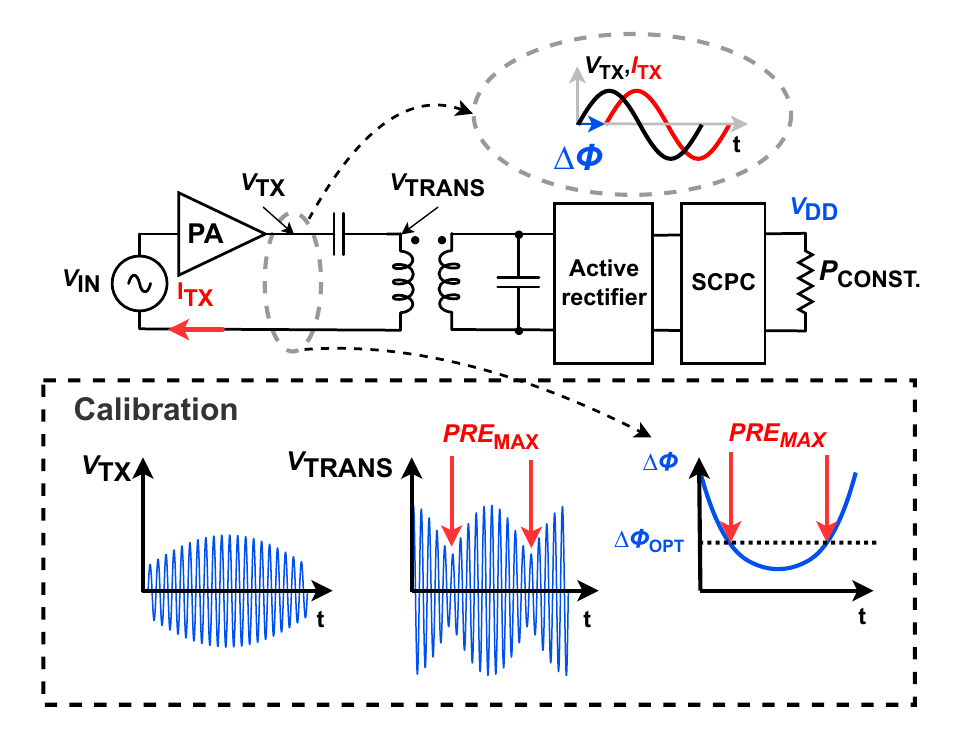}
    \caption{Illustration of the load sensing concept enabled by the proposed PMU implementation.}
    \label{fig:load_sensing}
\end{figure}

PRE optimization can be combined with a DC-DC converter after the rectification stage to transform any output load (R\textsubscript{L}) to an optimal \textit{R}\textsubscript{RX} by allowing \textit{V}\textsubscript{TX} to be decoupled from \textit{V}\textsubscript{DD}. In this scenario, global power control is necessary to regulate the TX power and maintain optimum PRE over output load variations. This work introduces a TX-side-only load sensing for closed-loop regulation requiring only a phase measurement and a single calibration step (\Cref{fig:load_sensing}), similar to \cite{Bai_2023}. At startup, the implant is programmed to draw constant power (nominal light load), while the TX AC voltage (\textit{V}\textsubscript{TX}) is swept over a predetermined range. The peak PRE is achieved when the \textit{V}\textsubscript{TRANS} amplitude is minimal, since \textit{V}\textsubscript{TRANS} is approximately proportional to the radiated power. Notably, the phase difference between \textit{V}\textsubscript{TX} and \textit{I}\textsubscript{TX} is proportional to \textit{R}\textsubscript{RX} and can be used to track PRE. At startup, the phase difference for maximum PRE is measured (\textDelta \textPhi\textsubscript{OPT}). During normal operation, the TX power is regulated to maintain \textDelta \textPhi\textsubscript{OPT}.

\begin{figure*}[!t]
    \centering
\includegraphics[width=0.99\linewidth]{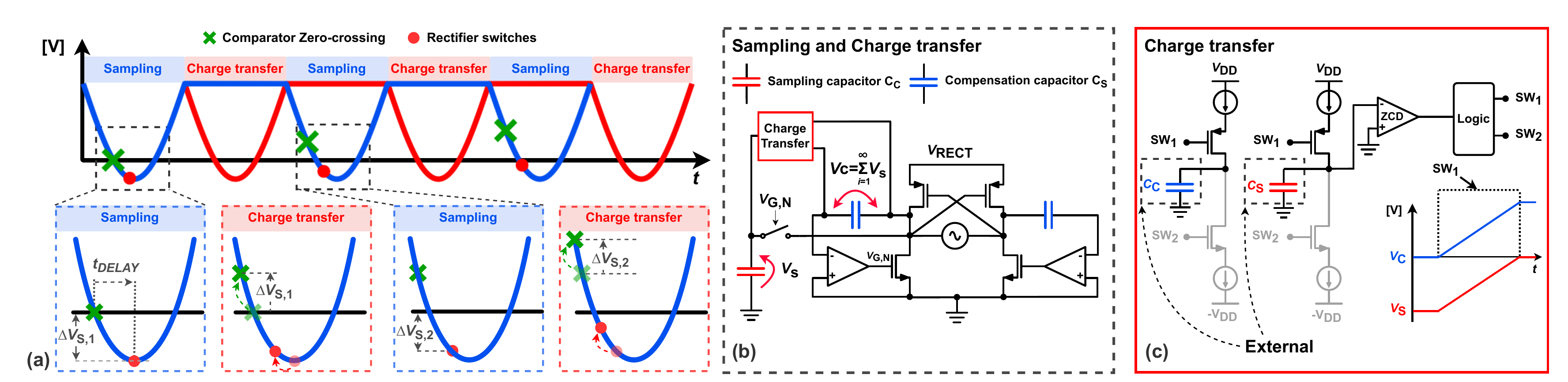}
    \caption{(a) The proposed delay compensation, (b) simplified operation of the sample-and-accumalate and (c) transient waveform of the voltage compensation.}
    \label{fig:Compensation_mechanism}
\end{figure*}

\section{Active Rectifier with Delay Compensation}

The delay compensation technique for the active rectifier in this work does not require a slow feedback control loop, as in previous works, to maintain system stability. Instead, it samples the error voltage ($\Delta V_S$) at the switching instant and adds it in series to the input of the comparator to compensate for the delay directly (\Cref{fig:Compensation_mechanism}(a)). A few accumulation cycles are typically sufficient to account for the nonlinear relationship between the error voltage and the required comparator offset to fully compensate the delay (simulated worst-case settling \textless 200~ns). After system startup, this technique can continuously compensate for the delay in a single cycle (25~ns) over typical load and input variations (120~\textOmega~-~1k\textOmega).

\begin{figure}
    \centering
    \includegraphics[width=1\linewidth]{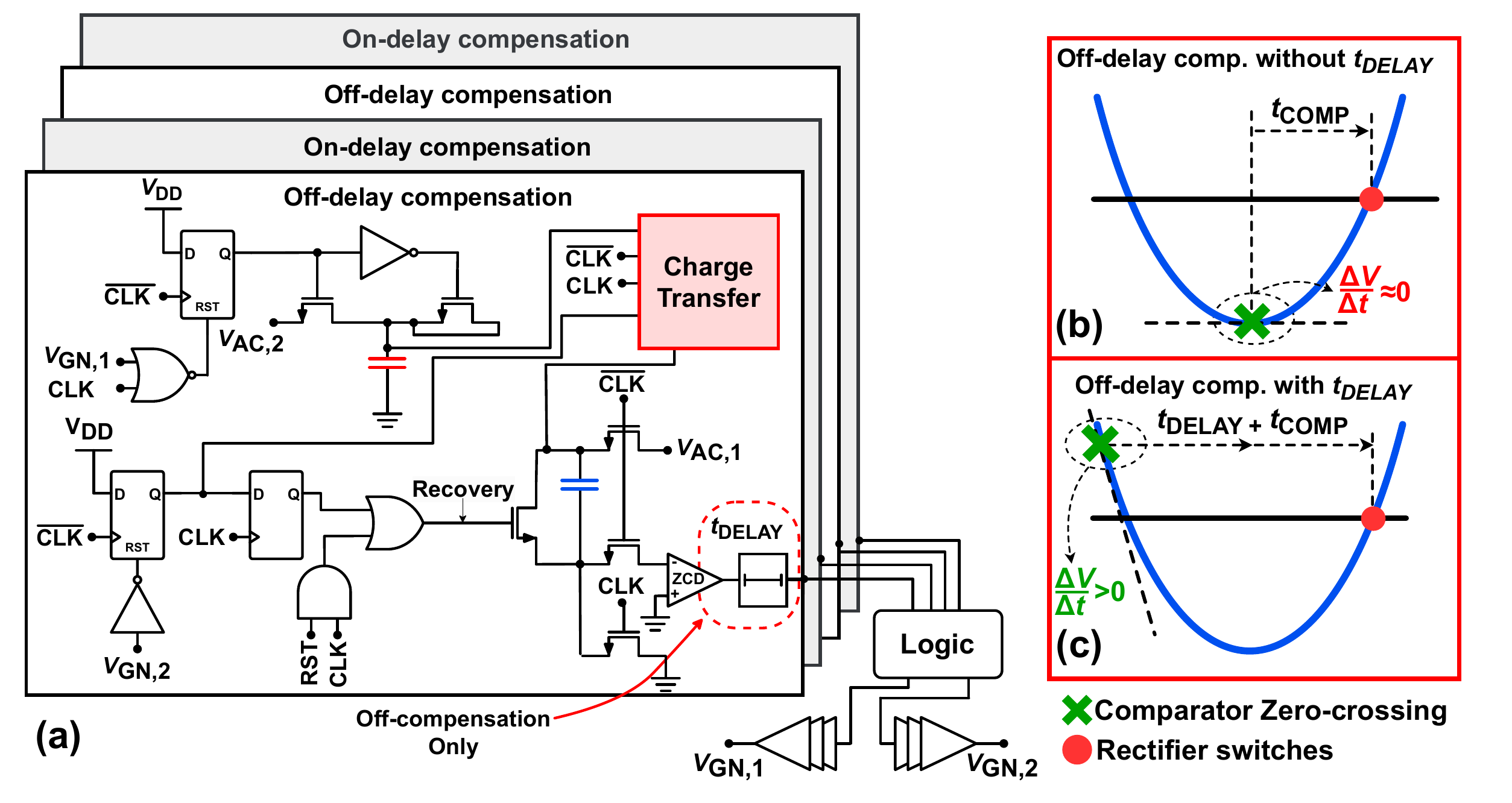}
    \caption{(a) Schematic of the delay compensation loops, (b) Off-delay compensation without delay and (c) Off-delay compensation with delay.}
    \label{fig:schematic1}
\end{figure}

The delay compensation implementation is shown in \Cref{fig:Compensation_mechanism}(b-c), where each edge of the gate driving signals (\textit{V}\textsubscript{GN,1},\textit{V}\textsubscript{GN,2}) has an independent compensation implementation. A sampling capacitor (C\textsubscript{S}) captures the error voltage at the switching instance of the rectifier, which is then transferred to a compensation capacitor (C\textsubscript{C}). A charge transfer circuit implements the sample-and-accumulate operation needed for the direct compensation technique using identical sampling and compensation capacitors. First, the polarity of the error voltage is determined to enable either the positive or negative charge-transfer current sources. Both capacitors are charged towards V\textsubscript{DD}~(-V\textsubscript{DD}) in case of a negative (positive) voltage using nominally identical current sources. A zero-cross detector (ZCD) ends the charging period for both capacitors when the voltage in the sampling capacitor crosses 0~V. The result is that the charge in the sampling capacitor is added to the compensating capacitor. The capacitors have the same capacitance and then the corresponding voltages are \textit{V}\textsubscript{S} and \textit{V}\textsubscript{C} respectively as shown in \Cref{fig:schematic1}. An additional safety circuit resets the compensation capacitor if it has an initial charge that disrupts the normal operation of the rectifier (\textit{V}\textsubscript{G,N} is always 0). The ZCD and the comparator for the active diode in the rectifier are implemented using auto-zeroed inverters \cite{tenten_1988}. Level shifters are used to switch the MOSFETs between the capacitors and the \textit{-V}\textsubscript{DD} rail and a SCPC generates -\textit{V}\textsubscript{DD} on-chip. 

\begin{figure}
    \centering
    \includegraphics[clip, trim=0.5cm 0.5cm 0cm 0cm,width=1\linewidth]{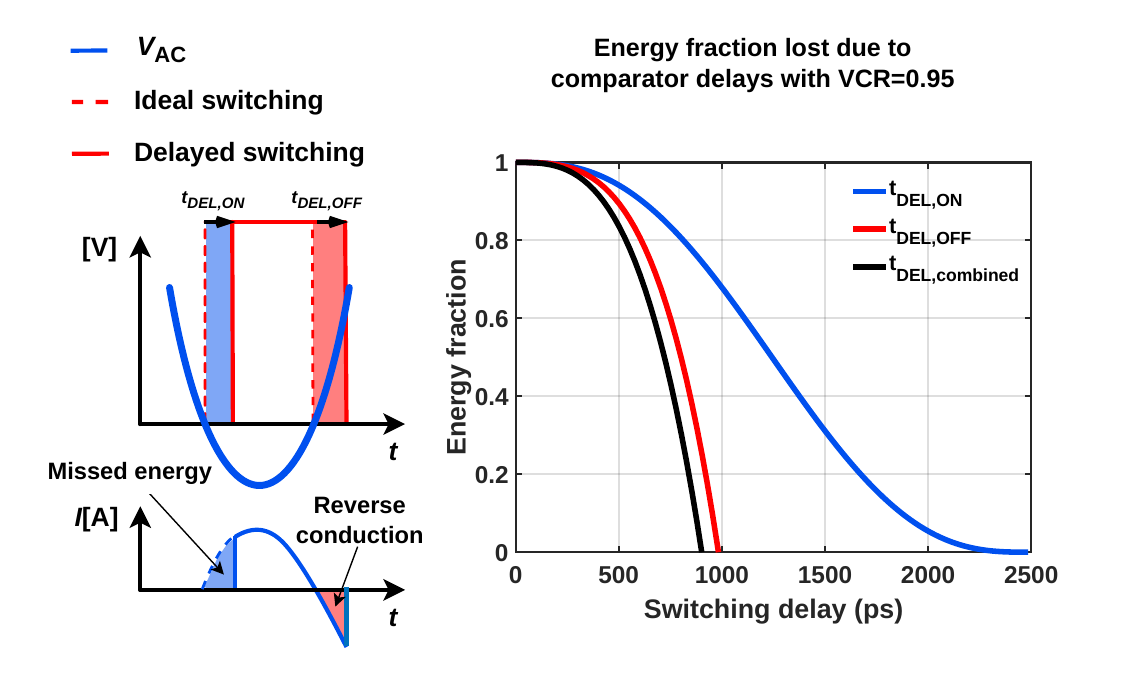}
    \caption{Impact of comparator delay on rectifier efficiency}
    \label{fig:delay_impact}
\end{figure}

Setting identical compensation architectures for both ON- and OFF-delays would move the OFF-trigger point close to the trough of the AC input, where $\frac{\Delta V}{\Delta t}$ approaches zero, making the system sensitive to voltage errors (\Cref{fig:schematic1}(b)). To address this issue, the OFF-compensation scheme adds a delay line, effectively shifting the OFF-trigger point to the same region as the ON-trigger point (\Cref{fig:schematic1}(c)). The compensation settles with a small residual delay \textless 200~ps, which is equivalent to \textless1\% energy losses due to ON/OFF delay (\Cref{fig:delay_impact}). 

\section{Measurement Results}\label{sec-meas}
\begin{figure}
    \centering
    \includegraphics[clip, trim=0.5cm 0.5cm 0cm 0cm,width=1\linewidth]{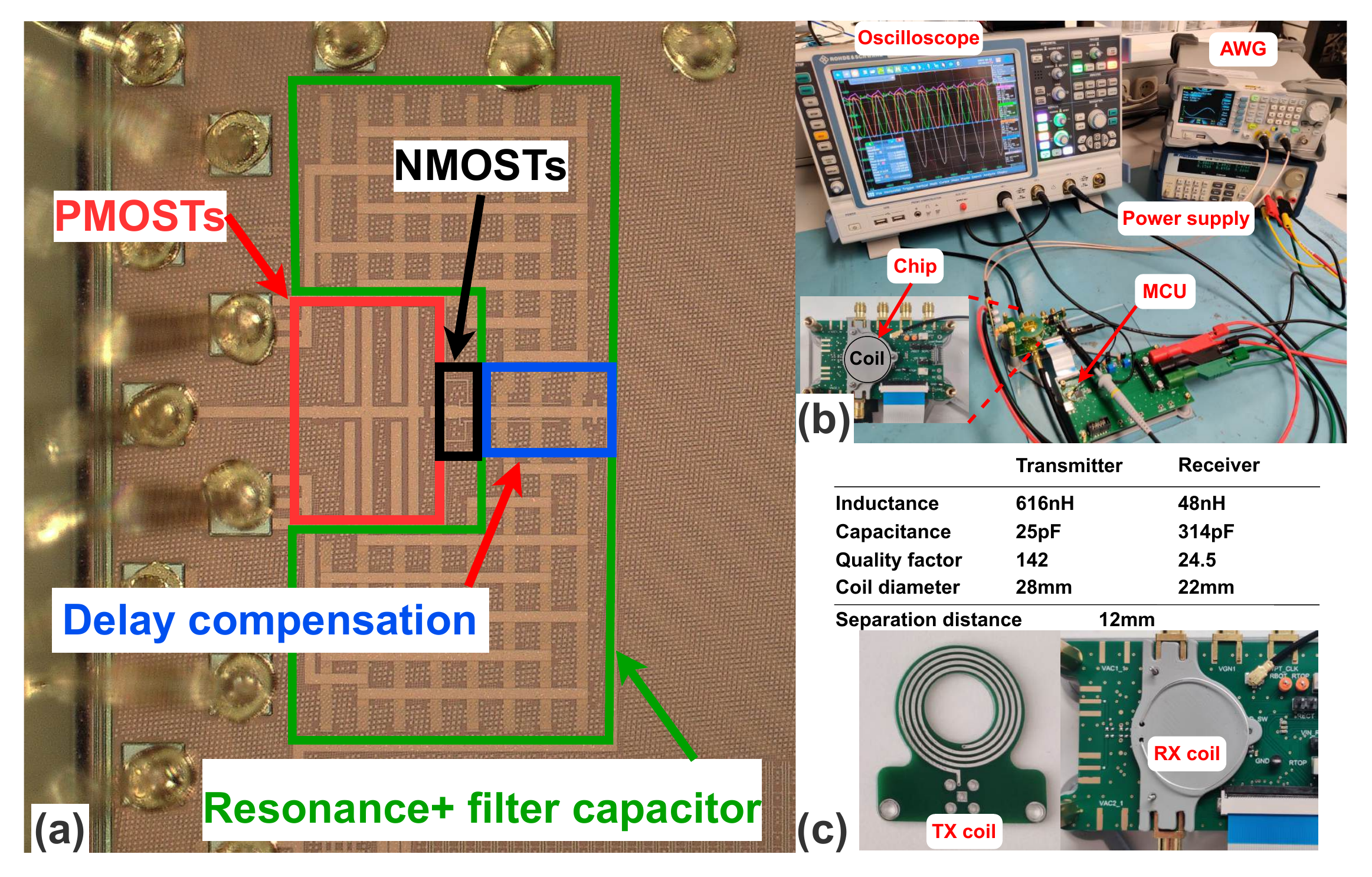}
    \caption{(a) Measurement setup in steady state and (b) the measured link parameters.}
    \label{fig:meas_setup}
\end{figure}
\begin{figure}
    \centering
    \includegraphics[clip, trim=0.5cm 0.5cm 0cm 0cm,width=1\linewidth]{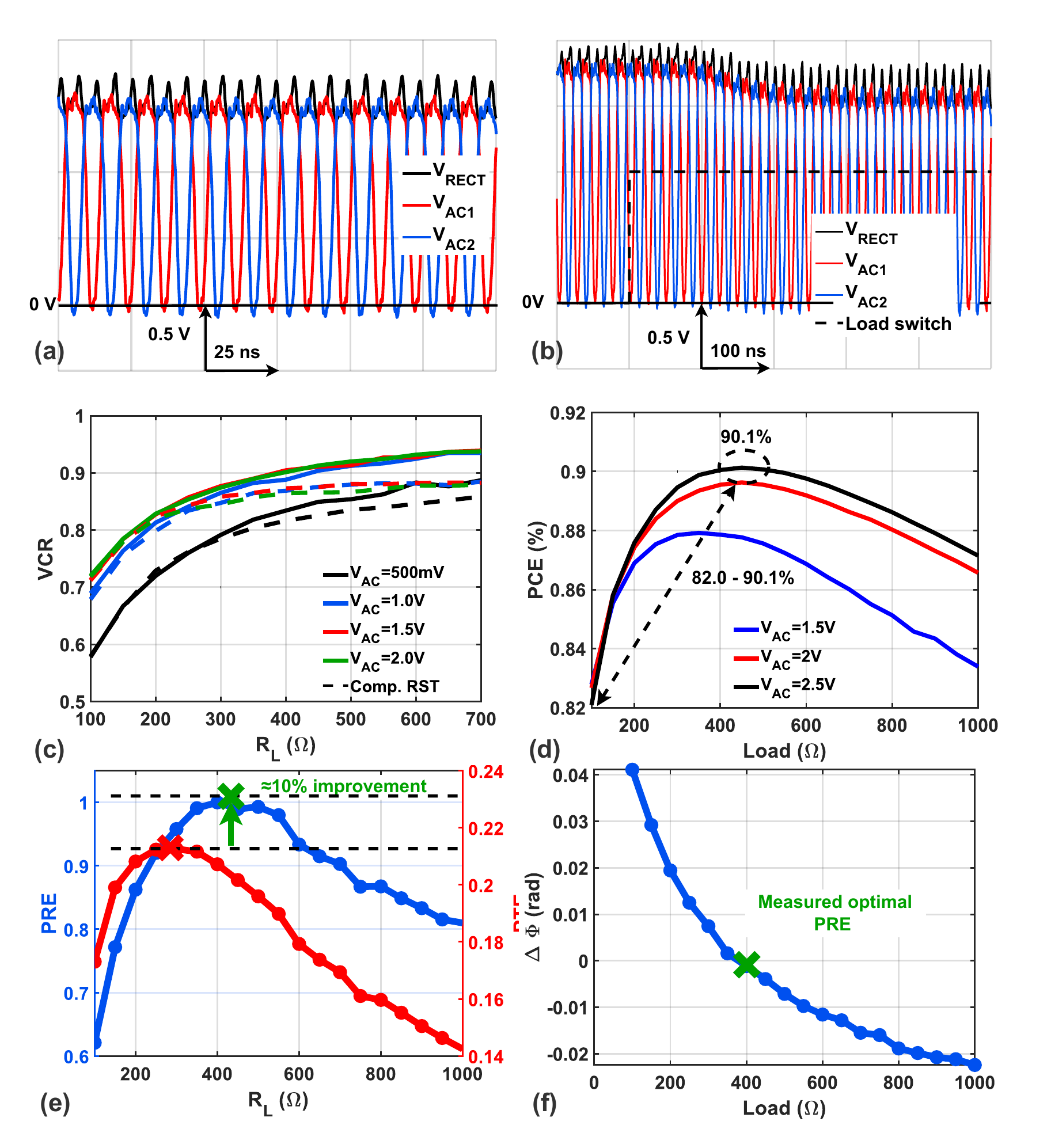}
    \caption{Measurement results for, (a) steady state transient, (b) stepped load from 600 $\Omega$ to 300 $\Omega$ transient, (c) VCR vs \textit{R}\textsubscript{L}, (d) PRE and PTE vs \textit{R}\textsubscript{L}, (e) phase vs \textit{R}\textsubscript{L} and (f) simulated PCE vs \textit{R}\textsubscript{L}.}
    \label{fig:metric_measurements}
\end{figure}

The proposed design was fabricated in a 40~nm CMOS process, occupying a chip area of 0.19~mm\textsuperscript{2}. The measurement setup consists of a 616~nH PCB inductor and tunable capacitors driven by a power amplifier IC on the TX side, while an RX board containing the wire-bonded ASIC completes the link (\Cref{fig:meas_setup}). High-speed buffer ICs (LMH6559) were used to minimize parasitic loading and reduce interference from the inductive fields when measuring the AC signals in the link. The steady-state and settling behavior of the measured system are shown in \Cref{fig:metric_measurements}(a-b). 

Rapid settling of the output can be observed with a load change from 600~\textOmega{} to 300~\textOmega. The output pole formed by the capacitance of the on-chip filter and the load (RC~=~150~ns) limits the settling of the rectifier and is consistent with the measurement results. It can be concluded that the compensation loop is likely faster, or as fast, as the observed settling. The delay compensation circuitry consumes only 130~\textmu W. The measured voltage conversion ratio (VCR) shows a peak value of 93.9\% with a 700~$\Omega$ load (\Cref{fig:metric_measurements}(c)). The PCE could not be measured due to complexities introduced by the on-chip capacitor and the UART communication interface, which made single-ended probing and direct driving of the chip impossible. The simulated PCE is shown in \Cref{fig:metric_measurements}(f), with a peak efficiency of 90.1\%.

The measurement results show a 10\% improvement when optimizing for PRE, instead of PTE (\Cref{fig:metric_measurements}(d)). The difference with the approximation in \Cref{fig:Conventional_vs_proposed}(b) is due to the fact that the measurement results include all losses experienced by the link. Any additional unmodeled losses are included in the radiated power, shifting the PTE curve closer to the PRE curve as link losses increase. Finally, the measured phase difference between \textit{V}\textsubscript{TX} and \textit{I}\textsubscript{TX} shows the dependence on \textit{R}\textsubscript{RX} that can be used to regulate the TX power without the need for back communication from the implanted RX (\Cref{fig:metric_measurements}(e)). 

\section{Conclusion}
This work achieves VCR and PCE comparable to the state-of-the-art, with a high power density across a wide AC input range and occupies only 0.19~mm\textsuperscript{2} (\Cref{fig:comparison_table}). This performance is achieved while integrating the resonant capacitor on-chip and achieving the fastest delay compensation settling in literature. Additionally, we introduced a new link optimization approach that can reduce radiated power across any output load, thanks to a TX-side load sensing technique when a DC-DC converter is implemented in the power management unit. 
\begin{table}
    \centering
    \caption{Comparison with state-of-the-art designs.}
    \label{fig:comparison_table}
    \includegraphics[width=1\linewidth]{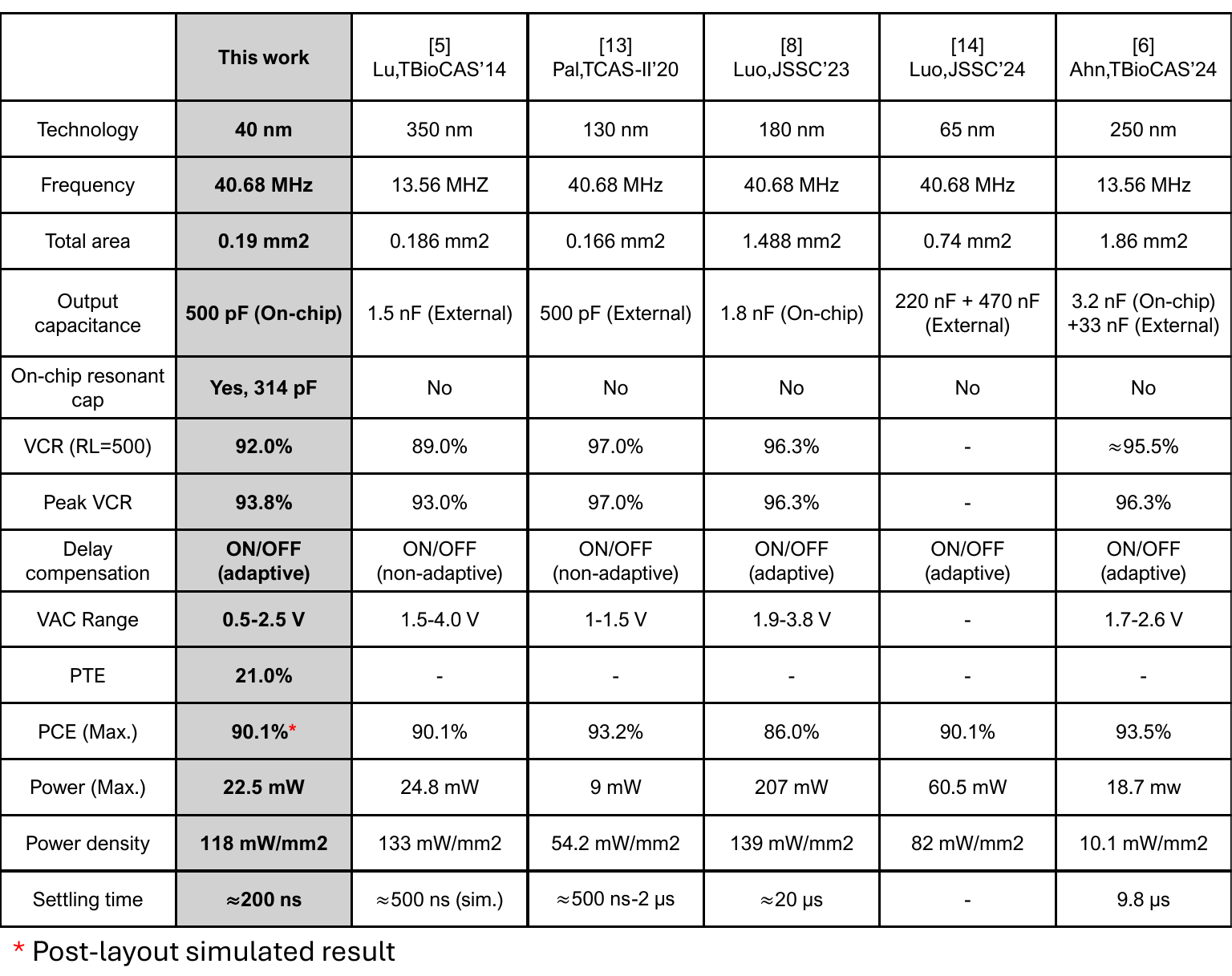}
\end{table}
\nocite{Pal_2020,Luo_2024}
\section*{Acknowledgments}
The authors thank Brian Nanhekan, Zu-Yao Chang, Marco Pelk and Juan Bueno Lopez for technical support.

\bibliographystyle{IEEEtran}
\bibliography{IEEEabrv,SSCL_paper}

\end{document}